\documentclass[iop]{emulateapj}
\shorttitle{10830 \AA\ Helium Line in M4}

\def\etal{{et al.}}

\usepackage{subfigure}
\usepackage{amsmath}
\usepackage{graphics}

\begin{document}

\title{The 10830 \AA\ Helium Line Among Evolved Stars in the Globular Cluster M4.}

\author{Jay Strader}
\affil{Department of Physics and Astronomy, Michigan State
University, East Lansing, MI 48824-2320}
\email{strader@pa.msu.edu}
 
\author{A. K. Dupree}
\affil{Harvard-Smithsonian Center for Astrophysics, Cambridge, 
MA 02138, USA}
\email{adupree@cfa.harvard.edu}
 
\author{Graeme H. Smith}
\affil{UCO/Lick Observatory, University of California, 1156 High St., 
Santa Cruz, CA 95064, USA}
\email{graeme@ucolick.org}

\begin{abstract}

Helium is a pivotal element in understanding multiple main sequences and extended horizontal branches observed in some globular clusters. Here we present a spectroscopic study of helium in the nearby globular cluster M4.  We have obtained spectra of the chromospheric \ion{He}{1} 10830 \AA\ line in 16 red horizontal branch, red giant branch, and asymptotic giant branch stars. Clear \ion{He}{1} absorption or emission is present in most of the stars. Effective temperature is the principal parameter that correlates with 10830 \AA\  line strength. Stars with $T_{\rm eff} < 4450$ K do not exhibit the helium line. Red horizontal branch stars, which are the hottest stars in our sample, all have strong \ion{He}{1} line absorption. A number of these stars show very broad 10830 \AA\  lines with shortward extensions indicating outflows as high as 80-100 km s$^{-1}$ and the possibility of mass loss. We have also derived [Na/Fe] and [Al/Fe] abundances to see whether these standard tracers of ``second generation" cluster stars are correlated with \ion{He}{1} line strength. Unlike the case for our previous study of $\omega$ Cen, no clear correlation is observed. This may be because the sample does not cover the full range of abundance variations found in M4, or simply because the physical conditions in the chromosphere, rather than the helium abundance, primarily determine the \ion{He}{1} 10830 \AA\ line strength. {A larger sample of high-quality \ion{He}{1} spectra of both ``first" and ``second" generation red giants within a narrow range of $T_{\rm eff}$ and luminosity is needed to test for the subtle spectroscopic variations in \ion{He}{1} expected in M4.}

\end{abstract}


\section{Introduction}

Messier 4 (M4) is one of the nearest globular clusters to the Sun, and as such, has received considerable attention via both photometry and  spectroscopy methods. The cluster has a horizontal branch that is well populated by stars both blueward and redward of the RR Lyrae region (Lee 1977). With a metallicity of ${\rm [Fe/H]} \sim -1.2$ (Ivans et al. 1999; Yong et al. 2008; Malavolta  et al. 2014) based on high resolution spectroscopy, M4 sits near the local minimum in the metallicity distribution of the Galactic globular cluster system that separates the two peaks corresponding to the metal-poor and metal-rich populations of clusters. Since the work of Norris (1981), who discovered a CN bimodality among the red giant branch stars, it has been known that M4 is chemically inhomogeneous. Abundance inhomogeneities have since been discovered among both red giant and horizontal branch stars in the elements ranging from carbon to aluminum (Ivans et al. 1999; Smith et al. 2005; Marino et al. 2008, 2011; Carretta et al. 2013).  The pattern of correlated Na--Al--CN and anticorrelated O--Na in M4 is also typical of many globular clusters  (e.g., Kraft 1994; Carretta et al. 2009a,b), and is often taken as evidence for multiple generations of star formation within the clusters (Gratton et al.~2012).

In M4 two sequences of main sequence stars have been detected using optical and near-infrared photometry (Milone et al. 2014; Nardiello et al. 2015). In the cluster $\omega$ Cen, Norris (2004) and {Piotto et al. (2005)} argued that the bluest of the main sequence components was enhanced in helium. Subsequent studies (cf.~Gratton \etal\ 2012) have concluded that helium is enhanced in other clusters too---perhaps as a result of high temperature hydrogen burning in a previous stellar generation. Spectral study of the 5875\AA\ \ion{He}{1} feature in the blue horizontal branch stars of M4 suggests they have a helium abundance of $Y=0.29$ which exceeds the primordial helium abundance by $\Delta Y \sim$ 0.04 (Villanova et al. 2012).

Given the relatively close proximity of M4 to the Sun, and the rather archetypal pattern of abundance inhomogeneities within it, this cluster seemed a natural additional to a program that we have been carrying out to study the behavior of the 10830 \AA\ \ion{He}{1} line among evolved Population II stars, both in clusters (Dupree et al. 2011; Smith et al. 2014) and in the Galactic halo field (Smith et al. 2004; Dupree et al. 2009). This paper reports the results of an investigation of a  sample of red giant branch (RGB), red horizontal branch (RHB), and asymptotic giant branch (AGB) stars in M4. Spectra of the 10830 \AA\ \ion{He}{1} feature for 16 stars have been obtained with the PHOENIX spectrograph on the Gemini-South telescope for a study of line profiles and equivalent widths, while high resolution optical spectra of the same stars with the Magellan/MIKE spectrograph have been used for a study of the Fe, Na and Al abundances to trace the abundance differences. Ancillary information is also available from the literature on the Na and Al abundances and 3883 \AA\ CN band strengths for a subset of the stars observed. 

While the main goal is to use the 10830 \AA\ \ion{He}{1} feature to provide abundance information (e.g., Pasquini et al. 2011; Dupree \& Avrett 2013), it is worth noting that this 
close triplet of lines formed in the upper chromosphere is very sensitive to mass motions, and indeed it was this context that the line was first studied in metal-poor giants (Dupree et al. 1992; Smith et al. 2004).

\section{Observations}

\subsection{Stars Observed in M4}

The sample of stars observed is listed in Table 1, designated by the nomenclature of Lee (1977). Photometry is listed in the Johnson $BV$ system as measured by Lee (1977), either through photoelectric or photographic techniques. Among the brightest stars a classification into RGB and AGB can be made by comparing the $V$ and $B-V$ photometry of each star with the $(V, B-V)$ color-magnitude diagram (CMD) of Cudworth \& Rees (1990; see their Figure 4) for stars with probabilities of cluster membership greater than 90\% on the basis of proper motion. There is differential reddening across M4 (Sturch 1977; Cudworth \& Rees 1990; Lyons et al. 1995; Hendricks et al. 2012), which leads to a spread in the color of the RGB and AGB sequences in the CMD, and complicates an unequivocal distinction between RGB and AGB stars, especially at high luminosities. The RGB and AGB sequences in the CMD are well-separated for $V>11.7$, and  the stars in our program are all fainter than this.  Smith \& Briley (2005) classified some stars in Table 1 on the basis of a CMD corrected for a single mean reddening. However, here we adopt the classification of Ivans et al. (1999) that considered differential reddening;  these classes are also listed in Table 1. 

\begin{deluxetable*}{lccccclcc}
\tablewidth{0pt}
\tablecaption{Data for Giants Observed in M4 \label{tab:basic}}
\tablehead{
ID\tablenotemark{a} & R.A. (J2000) & Dec. (J2000) &  $V$  &  $(B-V)$  & E(B-V)\tablenotemark{b}  &phot  & CMD  &  CN\tablenotemark{e} \\
                    & (deg)        & (deg)        & (mag) &   (mag) & (mag)  & type\tablenotemark{c} & class\tablenotemark{d} &  }
\startdata
L1404 & 245.820703 & $-26.559948$ & 13.40 & 0.94 &0.358& pg & RHB   &    \\
L1407 & 245.828622 & $-26.574282$ & 13.29 & 1.09 &0.385& pg & RHB   &    \\
L1408 & 245.819695 & $-26.574452$ & 11.79 & 1.41 &0.372& pg & AGB   &  S \\
L1617 & 245.875879 & $-26.556714$ & 12.18 & 1.39 &0.406& pg & AGB   &  S \\
L1701 & 245.874511 & $-26.530300$ & 12.02 & 1.38 &0.361& pg & AGB   &  W \\
L2413 & 245.823538 & $-26.452848$ & 13.38 & 0.96 &0.382& pg & RHB   &    \\
L2519 & 245.839155 & $-26.519267$ & 11.83 & 1.43 &0.363& pg & AGB   &  I \\
L3207 & 246.035779 & $-26.415329$ & 11.91 & 1.19 &0.300& pe & AGB   &  W \\
L3215 & 245.990519 & $-26.382156$ & 12.22 & 1.20 &0.310& pe & AGB   &  W \\
L3310 & 245.985505 & $-26.424570$ & 13.23 & 0.86 &0.319& pg & RHB   &    \\
L3503 & 245.967881 & $-26.506983$ & 13.19 & 0.98 &0.336& pg & RHB   &    \\
L4206 & 246.066187 & $-26.540953$ & 13.05 & 1.23 &0.310& pg & RGB   &    \\
L4406 & 245.950096 & $-26.612109$ & 13.36 & 0.93 &0.364& pg & RHB   &    \\
L4412 & 245.975814 & $-26.606150$ & 13.32 & 0.91 &0.354& pg & RHB   &    \\
L4413 & 245.993135 & $-26.594173$ & 12.62 & 1.30 &0.368& pg & RGB   &  S \\
L4415 & 245.973272 & $-26.570860$ & 12.52 & 1.36 &0.349& pg & RGB   &  S \\
\enddata
\tablenotetext{a}{ID from Lee (1977).}
\tablenotetext{b}{Reddening adopted in the analysis.  The line of sight towards M4 has a non-standard reddening law (Hendricks et al. 2012).}
\tablenotetext{c}{pe = photoelectric; pg = photographic.}
\tablenotetext{d}{Evolutionary status: RHB, RGB, or AGB. Taken from Ivans et al. (1999) for stars on the RGB/AGB.}
\tablenotetext{e}{CN class from Smith \& Briley (2005), with S, W, and I representing strong, weak, and intermediate.}
\end{deluxetable*}

A number of the most luminous RGB and AGB stars in the present program have been included in published studies of the CN distribution in M4. The classification of such stars into CN-strong, CN-intermediate, or CN-weak categories by Smith \& Briley (2005) is given in Table 1.  

\subsection{Gemini-S/PHOENIX Infrared Spectra}

Classical-mode observing time was awarded (Program GS-2009A-C-9) with the PHOENIX\footnote{See http://www.noao.edu/ngsc/phoenix/phoenix.html} spectrograph mounted on the Gemini-S telescope (Hinkle \etal\ 2003) to obtain high-resolution spectra of the \ion{He}{1} 10830\AA\ line of evolved stars in M4. Spectra were acquired over the course of three clear nights beginning 10 June 2009 UT. PHOENIX was configured with a slit of width 4 pixels yielding a spectral resolution of $\sim$50,000. The J9232 order-sorting filter was selected which spans 1.077--1.089 $\mu$m, and allows coverage of the \ion{He}{1} 1.083 $\mu$m line. Spectra of stars in M4 were acquired via a nodding mode (A-B pairs) with a spatial separation of 5 arcsec. Exposure times for the majority of the M4 giants totaled 3000 sec via two 1500 sec exposures, while for the fainter cluster stars observed the integrations totaled 6000 sec, and the brighter stars totaling 1800 sec via two 900 sec exposures. Because standard comparison lamps have a sparse wavelength pattern in this near-infrared region, a bright Population I K giant ($\alpha$ TrA) was also observed, the spectrum of which contains many securely identified narrow absorption lines from which to determine the wavelength scale. Throughout the night, spectra of rapidly-rotating B stars were obtained at a range of air masses to monitor the strength of telluric water vapor features. Additional standard stars, as well as the B stars, were observed with a nodding pattern having a separation of 8 arcsec, in order to avoid contamination of the spectra of the M4 giants by any residual image resulting from the observation  of these bright calibration stars. Flat-field lamp exposures and dark frames were taken at the end of each night.

The raw images were corrected for cosmic rays using L.A.~Cosmic (van Dokkum 2001). After this, the nodded image pairs were subtracted from each other. These subtracted images were divided by a dark-corrected, normalized flat field, followed by optimal extraction of one-dimensional spectra.  The spectra were wavelength calibrated using the observations of $\alpha$ TrA. The individual  wavelength-calibrated spectra were then combined with variance weighting. The summed spectra were normalized by fitting a second-order cubic spline to the continuum, excluding the region between 10833 to 10844 \AA\ where the Doppler-shifted \ion{Si}{1} line (10827.09 \AA) and \ion{He}{1} line (10830.34 \AA) occur. A narrow H$_2$O feature that sits 0.7\AA\ to 1.2\AA\ blueward of the strongest component of the \ion{He}{1} triplet was divided out using observations of fast-rotating hot stars taken at similar airmass. A mosaic of the observed 10830 \AA\ \ion{He}{1} profiles  is presented in Figure 1.

\begin{figure}
\figurenum{1}
\includegraphics[scale=0.53]{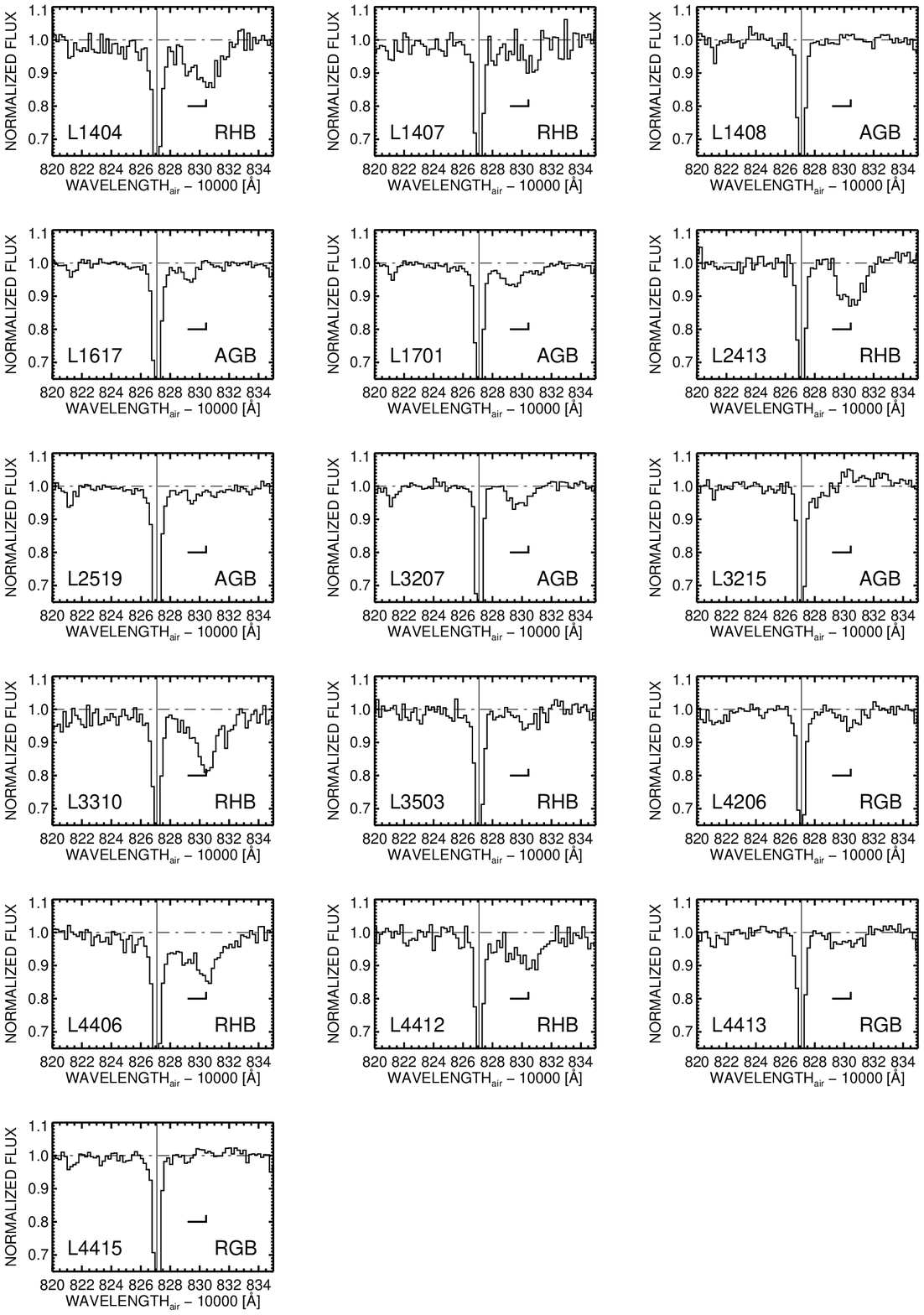}
\caption{Spectra in the vicinity of the 10830 \AA\ \ion{He}{1} feature for evolved stars in the globular cluster M4. The expected location of the \ion{He}{1} feature is marked in each panel, together with the designation of each star according to Lee (1977), and evolutionary state as proposed by Ivans et al. (1999).}
\end{figure}

Equivalent widths were measured by integrating the normalized spectra over the line feature in the IRAF task {\it splot}. Repeated independent measures suggest the typical random uncertainty in the equivalent width is  
$\sim 5$\%. {As an estimate of the systematic uncertainty, for the two RGB stars with detected \ion{He}{1} (L4206 and L4413), we remeasured the equivalent widths a number of times for different choices of the wavelength
range of the fitting and the continuum normalization, finding an uncertainty of about 15--20\%. This can be taken as a more conservative estimate of the uncertainty in the fitting procedure for \ion{He}{1} lines of
this equivalent width.}

Measurements of equivalent width of the \ion{He}{1} feature are listed in Table 2. In the case of two stars, L3215 and L3503, the listed value is a lower limit due to the possibility of emission redward of the anticipated rest frame center of the feature. Any such emission is admittedly subtle and subject to uncertainties in the continuum placement. Upper limits to the equivalent width are given for 3 stars (L1408, L2519, L4415) where no signature of helium could be found. The upper limits represent a direct integration over $\sim 1.5$ \AA\ centered on the wavelength of \ion{He}{1} in the target star determined relative to the strong \ion{Si}{1} absorption.

\begin{deluxetable*}{lrccccrcrr}
\tablewidth{0pt}\tablenum{2}
\tabletypesize{\scriptsize}
\tablecaption{Stellar Data and Abundances for Giants in M4 \label{tab:data}}
\tablehead{ 
ID\tablenotemark{a}  & $K_0$ & $(V-K)_0$  & $T_{\rm eff}$ & $\log g$ & [Fe/H] & 
[Na/Fe] & [Al/Fe] & He\tablenotemark{b} & EW(He)\tablenotemark{c} \\
                     & (mag) & (mag) & (K) & (cm s$^{-2}$) & (dex) & (dex) & (dex) & & (m\AA)} 
\startdata
L1404 & 10.44 & 1.63  & 5570 & 2.68 & $-1.21$ & 0.14    & 0.40 & yes & 321.7 \\ 
L1407 &  9.95 & 1.89  & 5226 & 2.45 & $-1.25$ & 0.03    & 0.67 & yes & 88.4 \\    
L1408 &  7.57 & 2.75  & 4394 & 1.40 & $-1.23$ & 0.16    & 0.53 & no  & $<20.7$ \\
L1617 &  8.09 & 2.54  & 4564 & 1.63 & $-1.28$ & 0.40    & 0.68 & yes & 54.4 \\
L1701 &  8.03 & 2.61  & 4503 & 1.60 & $-1.20$ & $-0.13$ & 0.44 & yes & 102.3 \\
L2413 & 10.29 & 1.66  & 5530 & 2.62 & $-1.27$ & 0.05    & 0.61 & yes & 232.4 \\  
L2519 &  7.62 & 2.84  & 4333 & 1.41 & $-1.04$ & $-0.27$ & 0.30 & no  & $<40.9$ \\
L3207 &  8.32 & 2.42  & 4646 & 1.73 & $-1.35$ & 0.06    & 0.56 & yes & 101.6 \\
L3215 &  8.68 & 2.37  & 4699 & 1.88 & $-1.14$ & $-0.22$ & 0.30 & em  & $>53.5$ \\
L3310 & 10.42 & 1.60  & 5616 & 2.68 & $-1.16$ & 0.11    & 0.57 & yes & 344.7 \\     
L3503 & 10.12 & 1.87  & 5252 & 2.52 & $-1.16$ & 0.01    & 0.41 & em  & $>68.9$ \\  
L4206 &  9.43 & 2.44  & 4623 & 2.17 & $-1.11$ & 0.41    & 0.65 & yes & 80.5 \\
L4406 & 10.30 & 1.70  & 5477 & 2.62 & $-1.24$ & 0.18    & 0.43 & yes & 357.4 \\     
L4412 & 10.40 & 1.61  & 5607 & 2.67 & $-1.20$ & 0.02    & 0.32 & yes & 225.2 \\      
L4413 &  8.79 & 2.46  & 4603 & 2.00 & $-1.22$ & 0.52    & 0.85 & yes & 88.5\\
L4415 &  8.56 & 2.70  & 4438 & 1.80 & $-1.10$ & 0.29    & 0.62 & no  & $< 10.7$ \\
\enddata
\tablenotetext{a}{ID from Lee (1977).}
\tablenotetext{b}{Indicates whether helium is present in absorption (yes), both emission and absorption (em), or not detectable (no).}
\tablenotetext{c}{The equivalent width of the helium line. For non-detections this is an upper limit; for the stars that show emission this is a lower limit.}
\end{deluxetable*}

\subsection{Magellan/MIKE spectra}

Echelle spectra of the  PHOENIX targets were obtained at the Magellan/Clay telescope on 25 May 2008 (UT), 14-15 July 2008 (UT), 3 May 2009 (UT), and 19 July 2013 (star Lee 4413 only), using the MIKE spectrograph (Bernstein \etal\ 2003) with a $0.7 \times 5\arcsec$ slit. The spectral resolution with this setup ranges from $\sim$26,000 on the blue side (3350--5000 \AA) to $\sim$36,000 on the red side (4900--9300 \AA). Total exposure times per star were typically 1800 sec in clear sub-arcsecond conditions. An IDL pipeline developed by S. Burles, R. Bernstein, and J. S. Prochaska\footnote{See http://www.lco.cl/telescopes-information/magellan/instruments/mike/idl-tools} was used to reduce the MIKE data. The pipeline evaluates the detector gain, processes a set of flats recorded each night, obtains wavelength solutions from exposures of a ThAr arc, constructs a slit flat-field frame, and extracts a spectrum including sky subtraction. The spectrum of the one star observed in 2013 was instead reduced with the Carnegie Python Pipeline ({\it CarPy})\footnote{See http://code.obs.carnegiescience.edu/mike}. Having produced wavelength-calibrated spectra for each MIKE echelle order, IRAF procedures were used to flatten (i.e., normalize to a continuum) the individual spectra, and then combine multiple exposures of each star. Absorption lines of \ion{Fe}{2}, \ion{Na}{1}, and \ion{Al}{1} were selected for abundance determinations from 8 echelle orders drawn from both blue and red sides of MIKE. Equivalent widths of each line were measured by using the IRAF routine {\it splot} to integrate directly over the normalized line profiles. In some cases, where blending was present, the blended profile was deconvolved using multiple Gaussian profiles in order to evaluate the contribution of the line of interest to the total feature. {The selected lines, adopted oscillator strengths, and measured equivalent widths are listed in Tables 3 and 4.}

\section{Abundance Analysis}

Fe, Na, and Al element abundances were derived from the MIKE equivalent widths using a procedure very similar to that described in Dupree \etal~(2011), which should be consulted for details. Minor modifications were made to account for  the large and differential reddening towards M4.

Initial estimates of effective temperature and surface gravity for the program star were obtained via a combination of broad band photometry. Specifically, $V$ and $K$ photometry (Table 2) was taken from the compilation of Hendricks \etal~(2012) for a subset of the stars. For the remainder that were not in this compilation, $V$ photometry was taken from Cudworth \& Rees (1990) and $K$ magnitudes from 2MASS (Skrutskie \etal~2006). Effective temperatures were determined from $(V-K)_0$ colors via the calibration of Alonso \etal~(1999).  These colors were corrected for differential reddening using the results of Hendricks \etal~(2012); extinction values for stars outside the boundaries of their map were obtained by linear extrapolation. {The fact that the $V-K$ colors compiled in Table 2 are in a slightly different photometric system to that used in the Alonso \etal~(1999) calibration was found to make a difference of less than 100 K in the inferred temperatures for the M4 stars in our MIKE sample, and we did not make a correction for this difference.} Surface gravities were estimated from the $K_0$ magnitude, assuming a stellar mass of $0.8 M_{\odot}$, and applying $K$-band bolometric corrections\footnote{While these are strictly defined only for $T_{\rm eff} < 5000$ K, for the warmer stars they agree with the relevant values from Bessell \etal~(1998) to within 0.01 mag.} from Buzzoni \etal~(2010). A distance of $1.80 \pm 0.05$ kpc was used (Hendricks \etal~2012).

The reddening is the largest source of random uncertainty in our analysis, as it can substantially change the derived effective temperatures. A change of 0.01 mag in $E(B-V)$ corresponds to a 35 K uncertainty in $T_{\rm eff}$. The effect on $\log g$ is smaller ($< 0.02$ dex) as the $K$ magnitudes are minimally affected by extinction and $T_{\rm eff}$ enters the surface gravity calculation logarithmically.

Model atmospheres were derived from the grid of Castelli \& Kurucz (2004) for [$\alpha$/Fe] = +0.4, interpolated to the temperatures and gravities of each star in the MIKE M4 sample. The values of $T_{\rm eff}$ and $\log g$ of each star were held fixed throughout the abundance analysis and  are listed in Table 2.  Lines used in the abundance analysis are given in Tables 3 and 4, along with measured equivalent widths. Measurements of the [Fe/H] abundance were carried out using an iterative fit of 10-13 \ion{Fe}{2} lines. The microturbulent velocity $v_t$ was estimated by requiring that there be no trend of Fe abundance with line equivalent width. Once these parameters were set, Na and Al abundances were determined from 1 to 5 clean absorption lines of each element. The stellar parameters and derived abundances are listed in Table 2.

{There are five stars in common between our sample and the larger M4 abundance survey of Ivans \etal~(1999)}, while there are four stars in common with Marino \etal~(2008). Table 5 provides a comparison between our assumed stellar parameters and derived chemical abundances to the previously published values. The $T_{\rm eff}$ scale adopted here is slight cooler than in previous studies, while the abundances generally appear to be consistent at the $\sim 0.1$ dex level.

\begin{deluxetable}{lrr}
\tablewidth{150pt}\tablenum{5}
\tablecaption{Comparisons to Previous Works\label{tab:data3}}
\tablehead{ 
Quantity  &  I99\tablenotemark{a}  &  M08\tablenotemark{b}}
\startdata
$T_{\rm eff}$(K)    &  $-71$    & $-97$   \\
$\log g$ (cm s$^{-2}$)        &  +0.10    & $-0.05$ \\
${\rm [Fe/H]}$ (dex)   &  +0.00    & $-0.04$ \\
${\rm [Na/Fe]}$ (dex)  &  $-0.08$  & $-0.06$ \\
${\rm [Al/Fe]}$ (dex)  &  $-0.17$  & $0.11$  \\ 
\enddata
\tablenotetext{a}{Median difference in parameters with Ivans \etal~(1999), in 
the sense of us--them.}
\tablenotetext{b}{Median difference in parameters with Marino \etal~(2008), in 
the sense of us--them.}

\end{deluxetable}

\section{Results}

\subsection{The GEMINI-S Spectra of the \ion{He}{1} Feature}

Figure 1 shows that considerable variety exists among the \ion{He}{1} profiles. Perhaps most striking is the strong \ion{He}{1} absorption of some of the RHB stars as compared to the RGB and AGB stars. L3503 has the weakest \ion{He}{1} feature among the M4 RHB stars in our sample, although even this star does appear to exhibit some He absorption. By contrast, the RHB stars L1404, L3310, L4406 and  L4412 all exhibit very strong absorption profiles that cover a wide wavelength range extending both blueward and redward of the rest wavelength. The profiles are broad, typical of a chromospheric line, in contrast to the nearby photospheric \ion{Si}{1} absorption feature. As the helium profiles are not typically symmetric about the rest wavelength, stellar rotation is not the cause of the broadening. For stars L1404 and L4406 in particular, the absorption appears to extend much more to the blue side of line center than to the red side. In these stars the absorption extends to the location of the \ion{Si}{1} line at 10827 \AA, implying mass outflows with speeds in excess of 80 km s$^{-1}$. Only in the instance of L3310 might it be argued that the \ion{He}{1} profile has broad but symmetrical wings that extend blueward to the vicinity of the \ion{Si}{1} line. By contrast, the only RHB star to have very strong \ion{He}{1} absorption that does not extend to the \ion{Si}{1} line is L2413. While all evolutionary states exhibit helium line absorption, more than half of RHB stars have strong, broad profiles with a notable degree of asymmetry. The one RHB star that most resembles the weaker \ion{He}{1} absorption seen among the RGB and AGB stars is L3503.

Figure 2 shows a color-magnitude diagram of our target stars with the presence of helium indicated. The absence of helium for the coolest stars in the sample is readily apparent.   

\begin{figure}
  \figurenum{2}
    \includegraphics[scale=0.51]{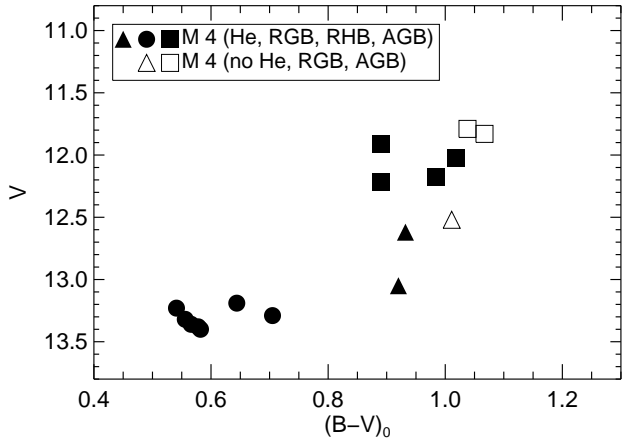}
    \caption{CMD for the observed stars in M4. The presence (or not) of helium absorption is indicated for various evolutionary stages.  Reddening was individually corrected based on the values from Table 1. Different symbols are used to highlight whether a star is considered to be on the RGB, RHB, or AGB. Filled symbols indicate helium is detected; open
symbols mark a non-detection of helium. }
  \end{figure}

Emphasizing this observation, Figures 3 and 4 shows the \ion{He}{1} plotted against $(B-V)_0$ and $T_{\rm eff}$, respectively. In Figure 3, all five RHB stars with $(B-V)_0 < 0.60$ have equivalent widths greater
than 200 m\AA, while all stars in our sample with $(B-V)_0 > 0.60$ (including the RHB stars L3503 and L1407) have equivalent widths less than 120 m\AA. Equivalent widths in the range 0-100 m\AA\ for the 10830 \AA\ \ion{He}{1} feature have also been found among the evolved giants of the globular cluster M13 (Smith et al. 2014), and metal-poor field stars in the halo (Dupree et al. 2009). Among RGB and AGB stars there is no clear trend between EW(He) and the $(B-V)_0$ color.

Figure 4 shows that the separation between stars having strong 10830 \AA\ features (EW $>  200$ m\AA) and those with EW $< 100$ m\AA\ occurs rather abruptly at $T_{\rm eff} \sim 5600$ K.  Helium is not detected in RGB and AGB stars with $T_{\rm eff} < 4450$K. 

These results may suggest that the hotter RHB stars have a greater far-UV flux from the chromosphere that ionizes helium and populates the lower level of the 10830 \AA\ transition through recombination. Or, instead, the warmer stars  may have a distinct chromospheric structure that happens to include higher temperatures. While the MIKE optical spectra of the \ion{Ca}{2} H and K lines of the RHB stars exhibit emission reversals near the line cores, a systematic correlation between this emission strength and the helium equivalent width is not apparent. We note that the near-IR and optical observations were not contemporaneous, and the potential for line variability cannot be ruled out. It is well known that helium-burning clump giants in the solar abundance Hyades cluster exhibit X-rays and appear to have a resurgent dynamo and magnetic activity cycles (Baliunas et al. 1983). 

\begin{figure}
\figurenum{3}
\includegraphics[scale=0.49]{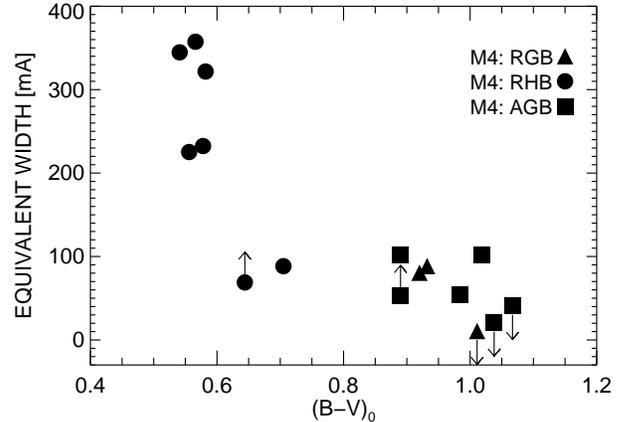}
\caption{The equivalent width of the 10830 \AA\ \ion{He}{1}  absorption feature versus $(B-V)_0$ color for evolved stars in M4. By far the strongest \ion{He}{1} lines occur among the RHB stars having colors of $(B-V)_0 < 0.60$, whereas among stars redder than this there is no clear trend between EW(He) and $(B-V)_0$.}
\end{figure}

\begin{figure}
\figurenum{4}
\includegraphics[scale=0.49]{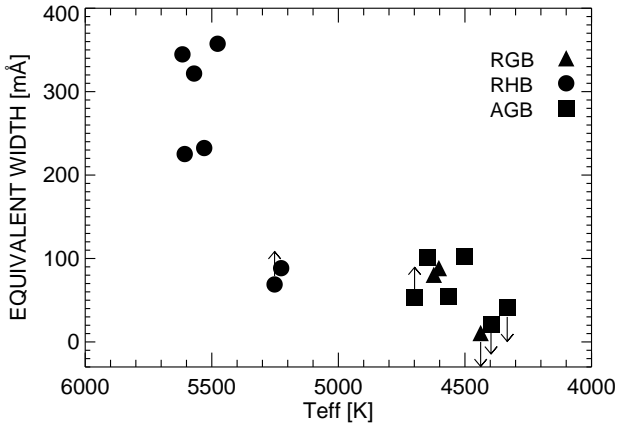}
\caption{Equivalent width of the \ion{He}{1} absorption feature versus effective temperature for evolved stars in M4.}
\end{figure}

L3207 and L3215, both of which are considered to be AGB stars, were included in a high resolution spectroscopic study of the NaD and H$\alpha$ lines by Kemp \& Bates (1995). They  found evidence for atmospheric mass motions in some M4 red giants, as exhibited by velocity shifts in the cores of the NaD and H$\alpha$ absorption profiles, although for  L3207 and L3215 any such shifts are on the order of 1 km s$^{-1}$ or less. Outflows generally accelerate through the chromospheres of red giants (M\'{e}sz\'{a}ros et al. 2009). L3207 exhibits a clear \ion{He}{1} absorption profile that is reasonable symmetric and does not exhibit high-velocity absorption, consistent with the NaD and H$\alpha$ information from Kemp \& Bates (1995). However, L3215 is a more difficult case to interpret, since there may be \ion{He}{1} absorption extending from the rest wavelength of the He line center towards the \ion{Si}{1} line. However, our spectrum of L3215 is rather unusual in the vicinity of any putative \ion{He}{1} feature since the spectrum appears to be high relative to the predicted continuum redward of line center. If real, this would be evidence of a P Cygni-type of line profile. This could instead be an artifact of an incorrectly set continuum redward of the \ion{Si}{1} line, though emission appears prior to the normalization of the spectrum, and other targets do not show a similar enhancement before they are normalized.

\subsection{The Magellan/MIKE Abundances}

The [Fe/H] measurements of this work can be compared with that of other prior investigations of M4. A mean value of [Fe/H] for the nine RGB or AGB stars in Table 1 is $-1.19$ with a range of 0.31 dex, while the seven RHB stars have a mean of $-1.21$ with a smaller range of 0.11 dex. There is no significant difference between these two means, and we conclude that our abundance analysis is consistent with ${\rm [Fe/H]} = -1.20$ for M4. This can be compared with means of $-1.18$ found by Ivans et al. (1999) from predominantly RGB stars, $-1.23$ by Yong et al. (2008), and $-1.16$ by Malavolta et al. (2014).

The helium equivalent width vs [Fe/H] is displayed in Figure 5 where no correlation appears between these two values.  This behavior is similar to that found both in field metal-poor giant stars (Dupree et al. 2009)  and
Omega Centauri giants (Dupree et al. 2011). We emphasize that we have not precisely estimated the uncertainties in the derived [Fe/H] values and we do not claim evidence for an intrinsic [Fe/H] spread in M4 on the basis of our spectra.

\begin{figure}
\figurenum{5}
\includegraphics[scale=0.49]{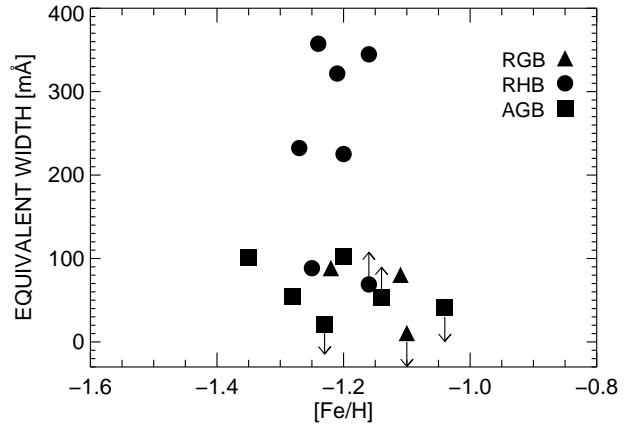}
\caption{Equivalent width of the \ion{He}{1} absorption feature versus [Fe/H] for evolved stars in M4.}
\end{figure}

The  trend between Na and Al abundances among the stars in our M4 sample is shown in Figure 6 as a plot of [Al/Fe] versus [Na/Fe], with different symbols being used to distinguish between RGB, RHB, and AGB stars using the classifications as given in Table 1. Although the number of RGB and AGB stars is small, there is nonetheless a clear correlation between [Na/Fe] and [Al/Fe] among these stars. Such an abundance pattern in M4 has been documented in much more detail for RGB stars by Ivans et al. (1999) and Carretta et al. (2013) and can be interpreted as spectroscopic evidence for multiple stellar populations in M4.

\begin{figure}
\figurenum{6}
\includegraphics[scale=0.49]{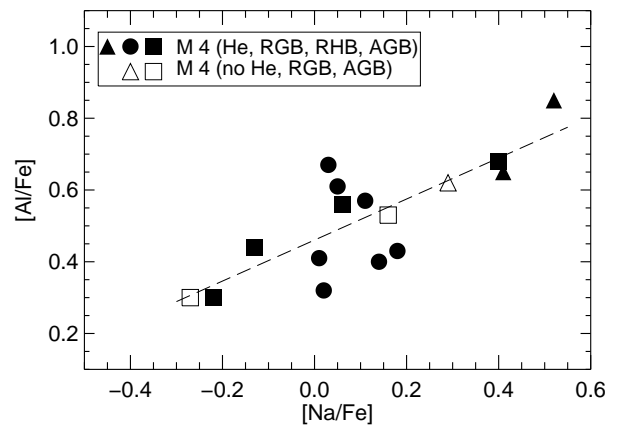}
\caption{The relationship between [Na/Fe] and [Al/Fe] abundances among the stars in the M4 sample of Table 1. Open symbols denote stars with upper limits (no detection) of helium. There is a Na--Al correlation evident among the RGB and AGB stars but not among the RHB stars.  The broken line represents a simple linear regression fit to the abundance values for the RGB and AGB stars.}
\end{figure}

The range in both [Na/Fe] and [Al/Fe] among the RHB stars in our sample is not large, on the order of 0.2 dex in [Na/Fe] and 0.4 dex in [Al/Fe]. There is no clear correlation in these abundances among the sample of RHB stars. This may be due to some extent to small sample statistics, since Marino et al. (2011) find a larger range in [Na/Fe] abundance among the RHB stars in M4 of 0.3--0.4 dex. Nonetheless, Marino et al. (2011) find that in M4 it is only when RHB stars are compared with blue horizontal branch (BHB) and RR Lyrae stars within the cluster that an anticorrelation between [Na/Fe] and [O/Fe] becomes apparent on the horizontal
branch. Thus the lack of a [Na/Fe]-[Al/Fe] correlation among the RHB stars in our M4 sample is not necessarily inconsistent with the trends between sodium and oxygen found along the entirety of the HB by Marino et al. (2011).

Based on the CN classifications in Smith \& Briley (2005), the CN-strong RGB stars in our sample are L1617, L4413, and L4415, with L2519 being classified as CN-intermediate, while the only CN-weak RGB star in our sample is L1701. For these five giants the [Na/Fe] abundances from Table 2 are 0.40, 0.52, and 0.29, $-0.27$, and $-0.13$. Thus the average [Na/Fe] for the three CN-strong giants is 0.40 dex. The CN-intermediate giant L2519 has a similar sodium abundance to the CN-weak giant L1701, and the average for the two stars is $-0.20$. Consequently our derived [Na/Fe] abundances are consistent with the CN-[Na/Fe] correlation in M4 discussed by Ivans et al. (1999) and Smith \& Briley (2005). The pair of CN-weak AGB stars, L3207 and L3215, have [Na/Fe] derived from the MIKE spectrum of 0.06 and $-0.22$ respectively, which for the case of L3215 is consistent with the sodium abundance of the CN-weak RGB star L1701. Star L1408 was listed by Smith \& Briley (2005) as having an uncertain evolutionary state, while the [Na/Fe] = 0.16 measured for it is midway between the CN-strong and CN-weak groupings.

\subsection{Comparing \ion{He}{1} Absorption with Na and Al Abundances and CN Strength}

Here we compare the \ion{He}{1} equivalent widths of the evolved M4 stars to the CN, Na, and Al abundances. Considering the CN-strong giants L1617, L4413, and L4415 the values of EW(He) are 54~m\AA, 88~m\AA, and $< 11$~m\AA. Thus two of these giants did exhibit \ion{He}{1} at the time of observation, while one did not. For the CN-intermediate giant L2519 the spectrum provided  only an upper limit on EW(He) of 40 m\AA. The CN-weak giant L1701 evinced a \ion{He}{1} feature of 100 m\AA\ that is greater than for any of the RGB stars with stronger CN absorption bands. Also interesting in this regard is that the two CN-weak AGB stars L3207 and L3215 have values of EW(He) of 102 and $> 54$ m\AA\ respectively, by contrast with $< 21$ m\AA\ for the CN-strong AGB star L1408. Thus in the case of both the limited number of RGB and AGB stars in our sample it is the CN-weak varieties that exhibit the strongest \ion{He}{1} absorption. To this can be coupled the observation that among the three CN-strong RGB stars L1617, L4413, and L4415 there is a range of some 70 m\AA\ in EW(He). Thus the data provide no evidence of any straightforward correlation between CN and He abundance among the evolved giants in M4. As such, our results for M4 are rather reminiscent of the findings of Smith et al. (2014) for the globular cluster M13.

Considering Na and Al, in Figures 7 and 8 we show plots of EW(He) versus [Na/Fe] and [Al/Fe] respectively. Here all the stars, including the RHB stars, are plotted. There are no clear trends in either of these plots. %

\begin{figure}
\figurenum{7}
\includegraphics[scale=0.49]{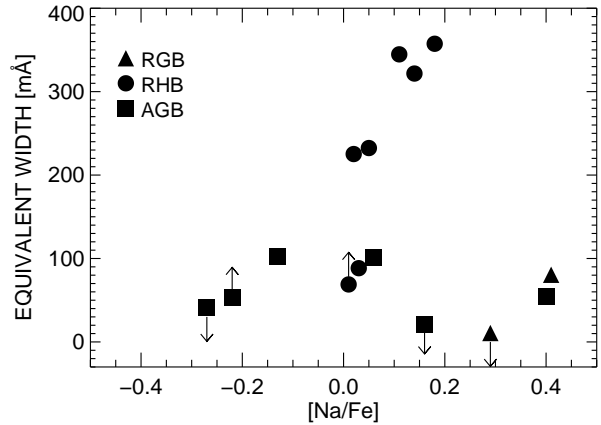}
\caption{Equivalent width of the \ion{He}{1} absorption feature versus [Na/Fe] abundance for evolved stars in M4.}
\end{figure}

\begin{figure}
\figurenum{8}
\includegraphics[scale=0.49]{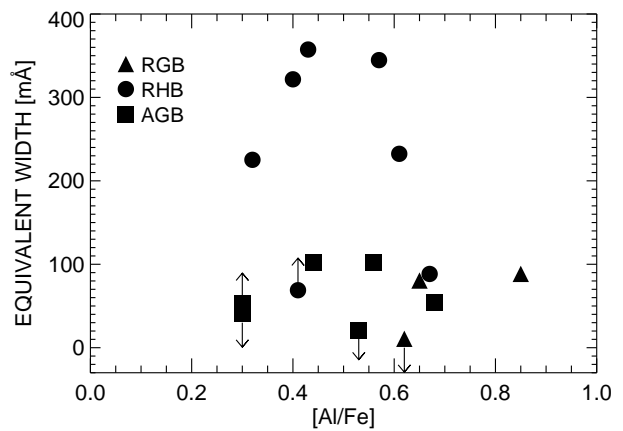}
\caption{Equivalent width of the \ion{He}{1} absorption feature versus [Al/Fe] for evolved stars in M4.}
\end{figure}

\section{Discussion}

We find no evidence for a correlation between the 10830 \AA\ \ion{He}{1} line strength and any abundance inhomogeneity in M4. As such, there is no evidence from our spectra that He abundance variations are implicated with the CN, O, Na, and Al abundance inhomogeneities that exist on the RGB and AGB of M4. 

The interpretation of these results is not straightforward. It may be that when the 10830 \AA\ feature is weaker than 120 m\AA\ it is much more sensitive to mass motions in the chromosphere or UV flux than it is to any potential range of He abundance that may exist in M4. Villanova et al. (2012) suggest that the helium enhancement in M4 over primordial values is small, $\Delta Y \sim$0.04. Indeed, the range of 10830 \AA\ He line strengths is much smaller in M4 than in the cluster $\omega$ Cen, in which this feature provides direct evidence of He abundance variations. (Dupree et al. 2011; Dupree \& Avrett 2013). {In M4 we have detected helium in two RGB stars with enhanced Na and Al abundances at an average equivalent width of 84 m\AA. Conservatively assuming that the expected change in the \ion{He}{1} equivalent width is comparable to the 15\% change in $\Delta Y$ inferred, then we predict that ``first generation" RGB stars should have 10830 \AA\ equivalent widths of about 72 m\AA. Given the typical measurement uncertainties, a sample of about 16 RGB stars of similar stellar parameters evenly divided between ``first" and ``second" generation stars would allow a $2\sigma$ detection of the predicted difference in helium between the two groups. This estimate is necessarily uncertain: at fixed $T_{\rm eff}$ a variety of effects can change the strength of the 10830 \AA\ line; on the other hand, in $\omega$ Cen we observed much larger differences in the \ion{He}{1} equivalent width than expected solely on the basis of a linear scaling of the line strength with the helium abundance.}

Arguably our most striking result is the marked increase in equivalent width of the 10830 \AA\ \ion{He}{1} line with increasing effective temperature on the horizontal branch of M4. This could be attributed to a real variation in He abundance from the red to the blue side of the RHB, or instead to a change in the structure of the chromospheres with increasing $T_{\rm eff}$. Determining whether He abundance variations are truly present would require careful, challenging modeling of the chromospheres of these stars, likely requiring additional data such as H$\alpha$ and \ion{Ca}{2} H and K spectroscopy (Dupree \& Avrett 2013). UV observations could also indicate if high temperature regions are present as a result of a magnetically heated atmosphere, providing a mechanism not only to strengthen the helium line but also driving the observed outflows.

One piece of circumstantial evidence consistent with true He abundance variations across the RHB in M4 is the result of Marino et al. (2011), who found a systematic difference in O and Na abundances between blue 
and red HB stars. The blue HB stars tended to have high Na and low O compared to the RHB stars: the Na--O anticorrelation found among RGB stars of comparable luminosity seems to map into a Na--O anticorrelation {\it along} the HB.  If the He abundance varies continuously with the Na abundance, then the warmer RHB stars might indeed be expected to have higher He, with this effect being masked in the RGB and AGB stars by other atmospheric effects. But at present we can only suggest this as a possible avenue for further spectroscopy and 10830 \AA\ line modeling.

The temperature cutoff of the appearance of helium for stars with $T_{\rm eff} <$ 4450K in M4, irrespective of evolutionary state, echoes a similar result found in the globular cluster M13 (Smith et al. 2004) and in  metal-poor field stars (Dupree et al. 2009). In this context, the claimed detection (Pasquini et al. 2011) of helium in one extremely cool luminous red giant in NGC 2808 ($T_{\rm eff}$= 3843K) is surprising and might be a transient phenomenon, perhaps arising during a phase of modest pulsation of the star. 

{We note that the $T_{\rm eff}$ limit below which helium is not observed in metal-poor stars is similar to that at which H$\alpha$ emission becomes prominent in RGB stars in the globular cluster M15, which was hypothesized to be related to the onset of pulsations in stars with $T_{\rm eff} \lesssim$ 4500K (Meszaros et al.~2008). This study also found that H$\alpha$ emission is present in AGB stars in M15 at warmer temperatures than for RGB stars, suggesting that both $T_{\rm eff}$ and evolutionary state are important determinants of the structure of the chromosphere.}

Less relevant for abundances, we also note the pronounced incidence of strong He absorption profiles among RHB stars in M4 and the frequency with which high-velocity absorption extends towards the \ion{Si}{1} line. Mass outflows with speeds of up to 80-100 km s$^{-1}$ are suggested by the blueward absorption profiles seen in some stars. The chromospheric escape velocity at 2 stellar radii equals $\sim 130$ km s$^{-1}$ for a star of mass 0.7 $M_{\odot}$ and $R= 8R_{\odot}$, higher than the velocities established for the RHB stars, though these values may be below the terminal velocities of the outflows. Strong \ion{He}{1} absorption and outflows have also been documented among RHB of the Galactic halo field population by Dupree et al. (2009), where speeds approaching chromospheric escape velocities are found. The results from M4 show that such outflows are also present in cluster RHB stars.

\acknowledgments

We thank an anonymous referee for comments that improved the paper. We thank Ben Hendricks for generously providing us with photometry and his differential reddening map of the field of M4.  We also thank Christian Johnson for facilitating  the Python reduction of a MIKE spectrum. This publication makes use of data products from the Two Micron All Sky Survey, which is a joint project of the University of Massachusetts and the Infrared Processing and Analysis Center/California Institute of Technology, funded by the National Aeronautics and Space Administration and the National Science Foundation. This paper is partially nased on observations obtained at the Gemini Observatory, which is operated by the Association of Universities for Research in Astronomy, Inc., under a cooperative agreement with the NSF on behalf of the Gemini partnership: the National Science Foundation (United States), the National Research Council (Canada), CONICYT (Chile), the Australian Research Council (Australia), Minist\'{e}rio da Ci\^{e}ncia, Tecnologia e Inova\c{c}\~{a}o (Brazil) and Ministerio de Ciencia, Tecnolog\'{i}a e Innovaci\'{o}n Productiva (Argentina).

{\it Facilities:}  \facility{Magellan:CLAY (MIKE spectrograph)}, \facility{Gemini-S (PHOENIX)}

\clearpage
\begin{turnpage}
\begin{deluxetable}{cccccccccccccc}
\def\a{\phantom{0}}
\def\b{\phantom{00}}
\tablecolumns{14}
\tabletypesize{\scriptsize}
\tablewidth{0pt}\tablenum{3}
\tablecaption{Atomic Parameters and Equivalent Widths for Fe\tablenotemark{a}}
\tablehead{
\colhead{Wavelength(\AA)}&
\colhead{4620.52}&
\colhead{4670.17}& 
\colhead{5100.66}& 
\colhead{5132.67}& 
\colhead{5197.58}& 
\colhead{5234.63}& 
\colhead{5264.79}& 
\colhead{5991.38}& 
\colhead{6084.10}& 
\colhead{6416.93}& 
\colhead{6432.68}& 
\colhead{6456.39}&
\colhead{6516.08}\\  
\colhead{Ion}  &     
\colhead{Fe II}&
\colhead{Fe II}&
\colhead{Fe II}&
\colhead{Fe II}&
\colhead{Fe II}&
\colhead{Fe II}&
\colhead{Fe II}&
\colhead{Fe II}&
\colhead{Fe II}&
\colhead{Fe II}&
\colhead{Fe II}&
\colhead{Fe II}&
\colhead{Fe II}\\
\colhead{log gf}&
\colhead{$-$3.23}&
\colhead{$-$4.05}&
\colhead{$-$4.20}&
\colhead{$-$4.01}&
\colhead{$-$2.24}&
\colhead{$-$2.10}&
\colhead{$-$3.02}&
\colhead{$-$3.57}&
\colhead{$-$3.81}&
\colhead{$-$2.68}&
\colhead{$-$3.57}&
\colhead{$-$2.13}&
\colhead{$-$3.28}\\
\cline{1-14}
\colhead{Star}&&&&\multicolumn{3}{c}{Equivalent Widths (m\AA)}&&&&&&&}
\startdata
L1404 & 52.2&16.9&11.1&23.2&111.7&114.5&50.1&23.0&20.9&35.2&36.2&66.6&48.5\\
L1407 & 51.8&25.4&21.9&14.8&102.6&91.9&41.8&31.3&13.5&30.8&33.9&61.0&45.9\\
L1408 & \nodata&21.1&23.9&18.5&98.1&95.9&36.7&32.8&21.9&31.5&42.8&59.2&59.7\\
L1617& \nodata&24.5&23.7&18.9&101.6\tablenotemark{b}&105.1\tablenotemark{b}&45.2&34.8&15.1:&30.6&42.1&64.8&62.2\\
L1701&58.3&26.4&21.8&20.4&95.7&95.1&44.8&30.5&21.3\tablenotemark{b}&30.2&42.2&59.8&56.7\\
L2413&\nodata&26.8&12.7&14.7&101.1&98.4&43.5&25.2&10.5&30.1&36.8&63.1&50.3\\
L2519&\nodata&29.9&23.4\tablenotemark{b}&20.3\tablenotemark{b}&92.6&96.3&44.5&31.7\tablenotemark{b}&21.9&27.8&42.8&54.7&59.1\\
L3207&\nodata&25.8&17.7&29.2&101.6&107.2&48.6&35.3\tablenotemark{b}&28.9\tablenotemark{b}&19.1&40.9&66.2&52.0\\
L3215&\nodata&23.9&21.8&26.7&99.2\tablenotemark{b}&100.1&50.3&31.2&18.7&32.9&45.6&63.8&61.5\\
L3310&\nodata&22.7&16.4&19.2&92.4&104.0&40.7&24.0&15.7&30.4&33.2&67.5&51.9\\
L3503&\nodata&24.9&13.7&19.6&99.2&105.3\tablenotemark{b}&45.9&28.9&19.0&37.8&39.5&67.5&57.5\\
L4206&42.6&22.9&\nodata&14.4\tablenotemark{b}&73.2\tablenotemark{b}&80.4&34.6&24.7&15.6&21.1&31.3&45.5&59.5\\
L4406&\nodata&22.9&19.4&29.7\tablenotemark{b}&14.1&100.8&107.3&43.8&15.4&20.1&39.1&65.7&48.3\\
L4412&\nodata&22.7&20.3\tablenotemark{b}&13.1\tablenotemark{b}&108.0&113.7&36.8\tablenotemark{b}&25.5&19.7&27.6&39.2&65.2&53.5\\
L4413&48.7&22.0&12.7&25.2:&69.8&73.6&33.9&23.1&11.4&18.3&30.6&48.6&46.2\\
L4415&47.5&21.2&19.6&17.9&75.8&77.1&27.5&23.0&13.6&23.1&32.8&43.9&41.7\\
\enddata
\tablenotetext{a}{Stars denoted by number from Lee (1977); Wavelengths and $gf$ values from Fulbright (2000). Equivalent widths given in m\AA.} 
\tablenotetext{b}{Feature was deblended.}
\end{deluxetable}
\end{turnpage}
\clearpage

\begin{turnpage}
\begin{deluxetable}{cccccccccc}
\def\a{\phantom{0}}
\def\b{\phantom{00}}
\tablecolumns{10}
\tabletypesize{\scriptsize}
\tablewidth{0pt}\tablenum{4}
\tablecaption{Atomic Parameters and Equivalent Widths for Na and Al \tablenotemark{a}}
\tablehead{
\colhead{Wavelength(\AA)}&
\colhead{4668.57}&
\colhead{5682.65}& 
\colhead{5688.21}& 
\colhead{6154.23}& 
\colhead{6160.75}&
\colhead{6696.03}& 
\colhead{6698.67}& 
\colhead{7835.32}& 
\colhead{7836.13}\\  
\colhead{Ion}  &     
\colhead{Na I}&
\colhead{Na I}&
\colhead{Na I}&
\colhead{Na I}&
\colhead{Na I}&
\colhead{Al I}&
\colhead{Al I}&
\colhead{Al I}&
\colhead{Al I}\\
\colhead{log gf}&
\colhead{$-$1.41}&
\colhead{$-$0.70}&
\colhead{$-$0.37}&
\colhead{$-$1.56}&
\colhead{$-$1.26}&
\colhead{$-$1.45}&
\colhead{$-$1.87}&
\colhead{$-$0.74}&
\colhead{$-$0.45}\\
\cline{1-10}\\
\colhead{Star}&&&\multicolumn{3}{c}{Equivalent Widths (m\AA)}&&&&}
\startdata
L1404 &5.4&32.8&45.4&\nodata&9.5&10.9&\nodata&\nodata&\nodata\\ 
L1407 & 4.5&27.9&47.6&6.7&7.6&25.7&10.2&17.9&24.6\\
L1408 & 23.1&84.5&106.9&23.1&41.3&42.8&23.0&30.3&39.1\\ 
L1617& 39.3&89.1&108.4&22.8&43.0&43.1&22.7&32.4&38.7\\
L1701&20.0&59.2&75.9&10.8&22.5&37.1&17.9&31.8:&29.8:\\
L2413&7.5&21.7&29.7&4.5&11.0&7.1&5.2&11.8&20.9\\
L2519&34.8:&70.0&87.3&13.8&29.7&45.3&22.9&28.5&35.7\\
L3207&12.0&50.4&66.7&15:&21.2&29.0&9.5:&22.8&24.9\\
L3215&4.6:&44.7&61.7&6.7&14.1&25.9&11.3&16.8&23.8\\
L3310&7.1:&31.5&39.9&2.8:&6.6&7.6:&\nodata&18.7:&15.7:\\
L3503&6.5&30.5&45.6&9.9&11.4&9.2&\nodata&12.4:&20.1\\
L4206&35.5&89.8&108.1&30.8&50.8&41.9&22.5&40.5&48.4\tablenotemark{b}\\
L4406&8.3&34.9&44.8&\nodata&6.4&17.3&\nodata&\nodata&11.0:\\
L4412&8.6&22.7&37.1&5.9:&11:&7.9&\nodata&19.6:&14.8\\
L4413&37.7&87.8&105.5&29.3&48.4&51.8&30.6&38.6&50.0\tablenotemark{b} \\
L4415&40.4&91.0&112.2&27.1&45.3&46.1&35.5&40.4&45.8\tablenotemark{b}\\
\enddata
\tablenotetext{a}{Stars denoted by number from Lee (1977); Na I wavelengths and $gf$ values from Fulbright (2000); those for or Al I from Jacobson et al. (2009). Equivalent widths given in m\AA.} 
\tablenotetext{b}{Feature was deblended.}
\end{deluxetable}
\end{turnpage}
\clearpage

\end{document}